\documentclass[pra,reprint,bibnotes,floatfix]{revtex4-1}

\usepackage[T1]{fontenc}
\usepackage[utf8]{inputenc} 
\usepackage[australian]{babel}

\usepackage[a4paper,centering,hmargin=1.75cm,vmargin=2cm]{geometry} 

\usepackage{amsmath,amssymb,graphicx,bm,microtype} 
\usepackage[dvipsnames]{xcolor} 
\usepackage{booktabs} 
\usepackage{siunitx} 
\usepackage{braket} 
\usepackage[colorlinks,allcolors=blue!50!black]{hyperref} 
\usepackage[all]{hypcap} 
\usepackage{enumitem}

\usepackage{array}

\begin{document}

\title{Classification of Coherent Enhancements of Light-Harvesting Processes}

\author{Stefano Tomasi}
\affiliation{School of Chemistry and University of Sydney Nano Institute, University of Sydney, NSW 2006, Australia}

\author{Ivan Kassal}
\email[Email: ]{ivan.kassal@sydney.edu.au}
\affiliation{School of Chemistry and University of Sydney Nano Institute, University of Sydney, NSW 2006, Australia}


\begin{abstract}
Several kinds of coherence have recently been shown to affect the performance of light-harvesting systems, in some cases significantly improving their efficiency.
Here, we classify the possible mechanisms of coherent efficiency enhancements, based on the types of coherence that can characterise a light-harvesting system and the types of processes these coherences can affect.
We show that enhancements are possible only when coherences and dissipative effects are best described in different bases of states.
Our classification allows us to predict a previously unreported coherent enhancement mechanism, where coherence between delocalised eigenstates can be used to localise excitons away from dissipation, thus reducing recombination and increasing efficiency.
\end{abstract}

\maketitle

Recent reports of coherent effects in complex light-harvesting systems have led to suggestions that coherent effects may play functional roles~\cite{Scholes2017}.
In particular, the spectroscopic observation of long-lived dynamical coherences in photosynthetic pigment-protein complexes~\cite{Engel2007,Liu2007,Panitchayangkoon2010,Collini2010} has stimulated discussion on whether coherence can enhance the performance of photosynthetic processes and whether it could inspire new design principles for artificial light-harvesting devices.

Arguments for functional coherences are complicated by the unnatural conditions of spectroscopic experiments, i.e., molecular dynamics under excitation by pulsed lasers differs dramatically from that under natural sunlight~\cite{Jiang1991,Mancal2010,Brumer2012,Kassal2013,Brumer2018}. Therefore, although spectroscopy provides critical information on the nature of chromophores and their couplings, relating experimentally detected coherences to possible functional roles remains an outstanding challenge~\cite{Scholes2017}. Many theoretical studies have taken on the task, providing potential mechanisms by which various types of coherence can occur in natural light-harvesting systems and improve their function.

Here, we unify this literature by classifying the possible types of coherence and the ways they can improve a light-harvesting process.

We consider light-harvesting systems that are aggregates of coupled light-absorbing sites (such as molecules or chromophores), which transport excitations (excitons) to an acceptor that converts them to useful energy.
Excitons can be introduced into the system either by direct illumination or through excitonic energy transfer (EET) from other systems.
The subsequent exciton dynamics is mediated both by unitary evolution under the system Hamiltonian and by interactions with the environment~\cite{maykuhn,breuer,Levi2015}.
Ultimately, excitons are either successfully transferred to the acceptor or lost to recombination.

The performance of the light-harvesting system can be described by its quantum efficiency, the proportion of excitons that reach the acceptor, as opposed to being lost to recombination.
The efficiency results from a kinetic competition between system-to-acceptor transfer and recombination, and can hence be improved either by increasing donor-acceptor transfer rates (trapping) or by decreasing recombination rates~\cite{Cao2009}.

\begin{figure}[t]
    \centering
    \includegraphics[width=0.95\linewidth]{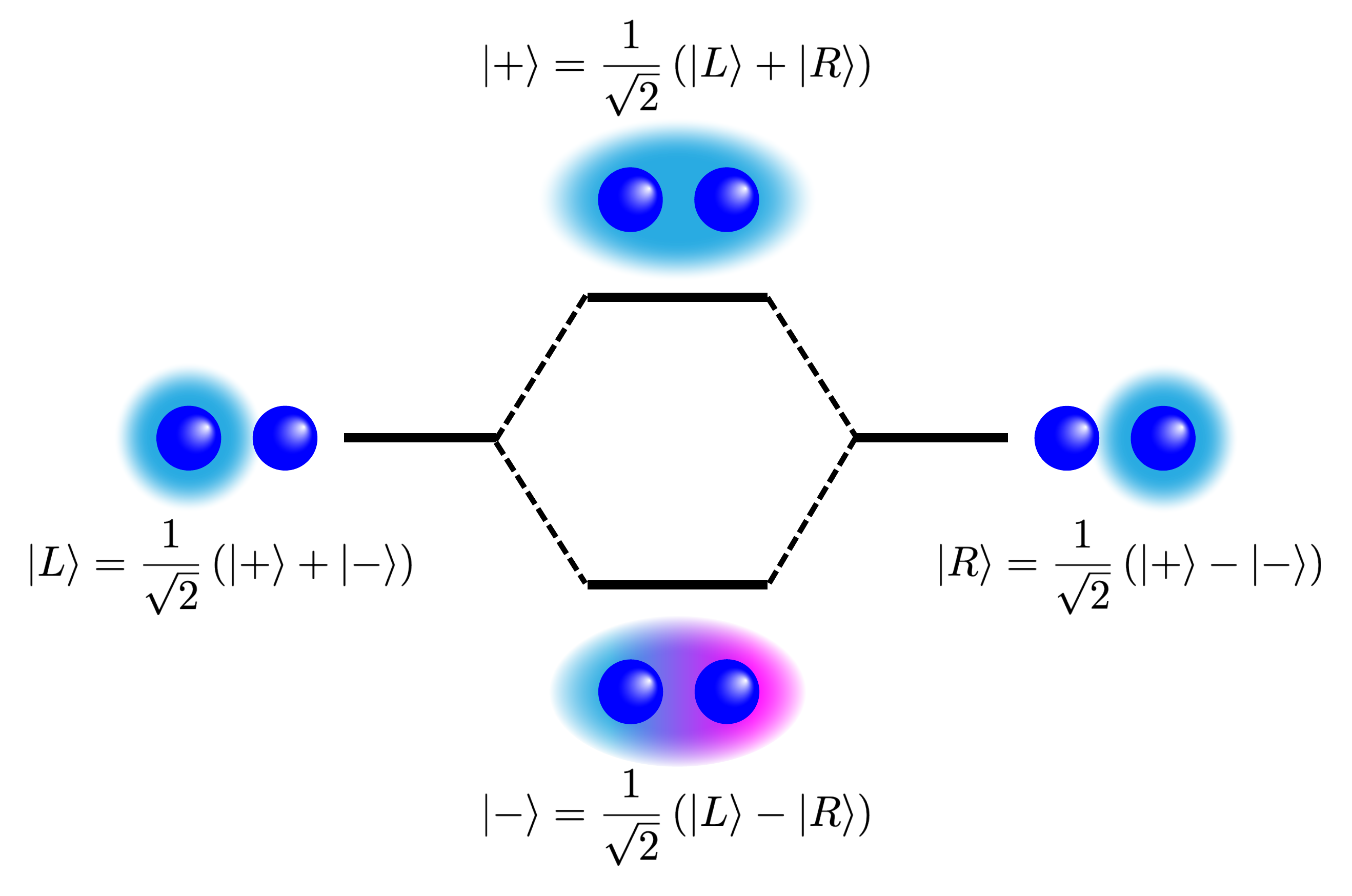}
    \caption{Site and energy bases in a two-site excitonic system. Coupling between the sites $\ket{L}$ and $\ket{R}$ causes the eigenstates $\ket{+}$ and $\ket{-}$ to be delocalised, i.e., coherent superpositions of site states. Localised site states, on the other hand, can be described as coherent superpositions of the two eigenstates.
    }
    \label{fig:1}
\end{figure}

We limit our discussion to enhancements in the single-exciton manifold because the weak light-matter coupling usually ensures that at most one exciton at a time is present in light-harvesting systems under natural illumination~\cite{Fassioli2012,Giusteri2016}.
Nevertheless, enhancements due to coherence in multi-exciton systems have also been proposed, and apply to systems characterised by strong light-matter interactions or long-lived excitons~\cite{Higgins2014,Higgins2017,Brown2019}.

Our classification of coherent enhancements is based on whether the trapping and recombination mechanisms act locally on individual sites or on states that are delocalised across multiple sites. Overall, we show that coherence can enhance efficiency only when the coherence occurs in a different basis to that in which the trapping or recombination act. We first survey the possibilities before addressing how the coherences might be generated.

\begin{table}
\centering
\renewcommand\arraystretch{1.3}
\begin{tabular}{>{\raggedright}p{1.5cm}>{\raggedright}p{3cm}>{\raggedright\arraybackslash}p{3cm}}
    \toprule
    & Site-basis coherence (delocalisation) & Energy-basis coherence  \\ \midrule
    
    Site-basis dissipation & 
    Enhancement impossible &
    Type II enhancements:
    \begin{itemize}[leftmargin=*]
        \item Trapping enhancement (Type IIA): Localisation near trap
        \item Recombination suppression (Type IIB): Localisation on quiet site
    \end{itemize}  \\
    
    Energy-basis dissipation & 
    Type I enhancements:
    \begin{itemize}[leftmargin=*]
        \item Trapping enhancement (Type IA): Supertransfer
        \item Recombination suppression (Type IB): Dark-state protection
    \end{itemize} & 
    Enhancement impossible \\
    \bottomrule
\end{tabular}
\caption{Whether or not a type of coherence can cause an enhancement of light-harvesting efficiency depends on the dissipation (trapping and recombination) mechanisms that affect it. These mechanisms can be classified based on whether they affect site populations or eigenstate populations. Efficiency enhancements (which can be either trapping enhancement or recombination suppression) are possible when at least one dissipation mechanism acts in a different basis to the coherence.
}
\label{t:class}
\end{table}

\section{Classifying coherences and trapping mechanisms}

Light-harvesting efficiency is a macroscopic property, the average of the successes and failures of many exciton trapping attempts. Therefore, it is best calculated using the ensemble density matrix $\rho$ of the system. The diagonal terms of $\rho$ represent \emph{populations}---the probabilities of the system being in certain states---its off-diagonal terms are \emph{coherences} and they quantify the extent to which the system is characterised by superpositions of quantum states.
Coherences are basis-dependent because $\rho$ may be diagonal in one basis but not in another~\cite{Kassal2013}. Because coherences in different bases can have different manifestations and consequences, it is important to specify the basis of the coherence.

Two bases are crucial for excitonic systems: the \emph{site basis} and the \emph{energy basis}~\cite{Kassal2013}.
The site basis is the set of states describing excitons localised on individual sites, while the energy basis comprises eigenstates of the system's Hamiltonian.
Therefore, coherence in the site basis is a characteristic of states that are delocalised across multiple sites, while coherence in the energy basis characterises superpositions between energy eigenstates.
When inter-site coupling is large compared to the energetic disorder, eigenstates tend to be delocalised, i.e., characterised by large site-basis coherence~\cite{maykuhn,Amerongen,Scholes1993,Dubin2006}.
Similarly, localised states can result from coherence between delocalised eigenstates due to constructive and destructive interference (Figure~\ref{fig:1}).

Light-harvesting efficiency is determined by the kinetic competition between two types of dissipation: recombination and trapping.
Recombination is the loss of excitons due to relaxation to the ground state (whether radiative or non-radiative), while we define trapping as the extraction of energy from the system through transfer to an acceptor state.
Trapping and recombination are, respectively, desirable and undesirable types of dissipation, and the efficiency of a light-harvesting system can be improved either through the enhancement of trapping or the suppression of recombination~\cite{Cao2009}.

Either dissipation process can, in principle, act in any basis, meaning that it depletes populations of states in that basis. However, most models of dissipation assume that it acts in the site or energy bases. 
For example, site-basis dissipation---which acts locally---can occur when a specific site in the system is coupled most strongly to an acceptor, while energy-basis dissipation---which acts non-locally---can occur in an ensemble of molecules collectively decaying into far-field radiation from a delocalised eigenstate. 
Here, we focus on these two types of dissipation, while noting that it is possible for dissipation to occur in another basis as well.

Whether coherence can affect light-harvesting efficiency depends on what basis the coherence is in and what bases the dissipation processes act in~\cite{Leon-Montiel2014,Tscherbul2018}.
An enhancement due to coherence is impossible when the coherence and all dissipation mechanisms are in the same basis. In that case, all dissipation processes depend only on the populations in the common basis, meaning that coherences in the same basis cannot affect either trapping or recombination (diagonal cells of Table~\ref{t:class}).
On the other hand, enhancements are possible when the coherence and at least one dissipative mechanism are in different bases (off-diagonal cells of Table~\ref{t:class}).
In those cases, a change in coherence in one basis affects populations in the other, and can therefore affect the outcome of dissipative process and the overall efficiency.

These conditions for coherent enhancement are analogous to known conditions for coherent control of observables, which state that for energy-basis coherence to affect measurement outcomes, the measured observable must not commute with the system Hamiltonian or the control must be assisted by an environment~\cite{Spanner2010}.
In our case, the measurement corresponds to a trapping or recombination event. The idea can be extended to site-basis coherence, which can affect the outcome of a measurement if the corresponding observable does not commute with a site projection operator.

These observations allow us to classify types of coherent enhancement based on the bases of the coherence and the dissipative mechanisms.
First, we define two types of coherent enhancement, summarised in Table~\ref{t:class}: type I enhancements, due to site-basis coherence, and type II enhancements, due to energy-basis coherence. 
This dichotomy is most productive in the most usual cases, noted above, where dissipation acts in one of these two bases; otherwise, if the dissipation acted in another basis, the relevant consideration, as before, would be whether it was in the same basis as the coherence.
Second, we define two sub-types within each I/II category, depending on how the efficiency is increased: types IA and IIA are due to trapping enhancement, and types IB and IIB are due to recombination suppression.
These two cases are the limits of a continuous spectrum, because in larger systems with many degrees of freedom, a change in dynamics due to coherence may not unambiguously lead to only the suppression of recombination or the enhancement of trapping, but instead to an overall change in efficiency that results from a combination of both. Nevertheless, the A/B distinction is useful for gaining intuition about the possible varieties of coherent enhancement.

\section{Type I enhancements}

\begin{figure}[tb]
    \centering
    \includegraphics[width=\linewidth]{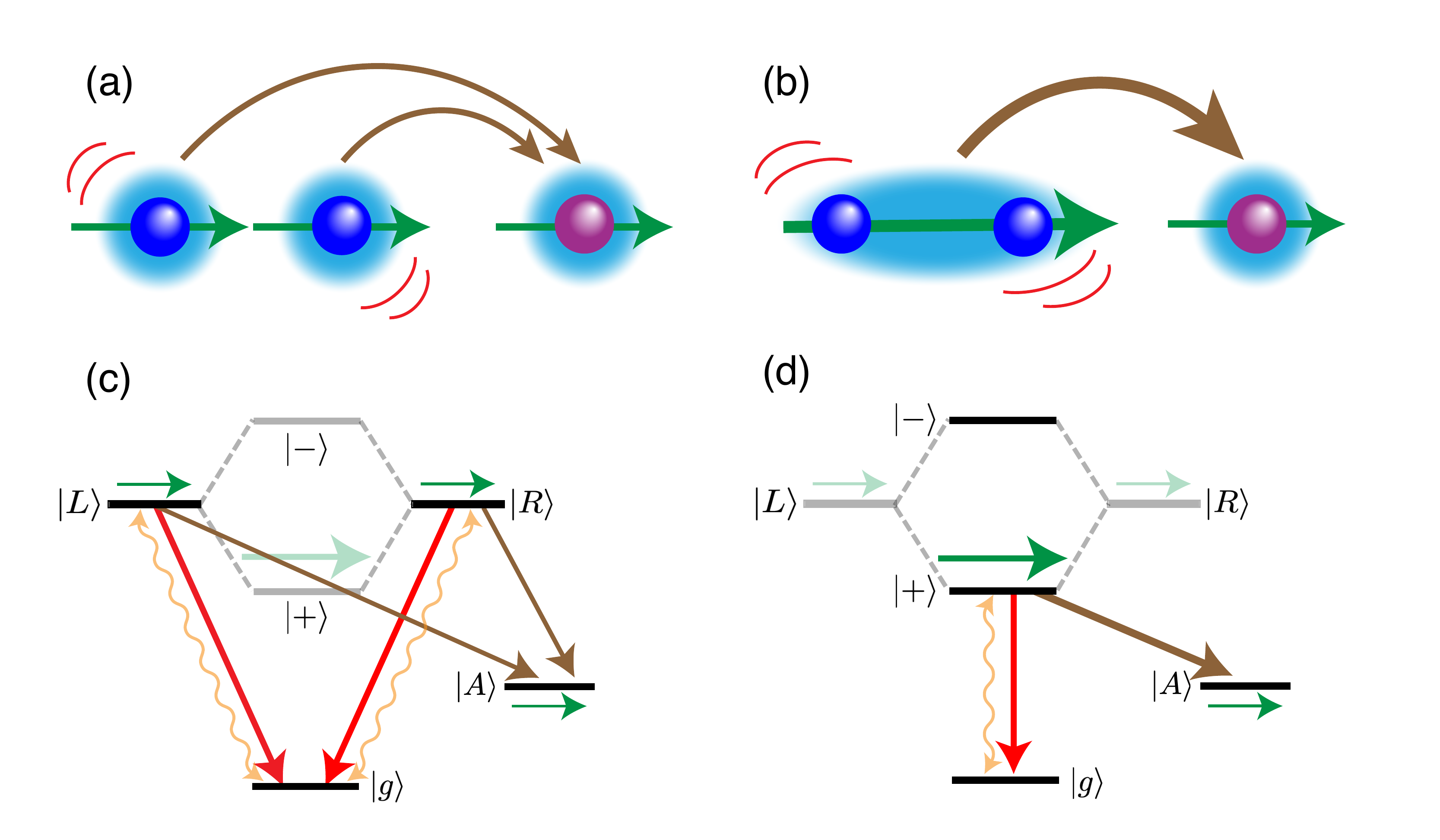}
    \caption{Type IA enhancement via supertransfer.
    A system comprising two identical sites ($\ket{L}$ and $\ket{R}$, blue) and a redshifted acceptor ($\ket{A}$, purple), either in an incoherent mixture of donor sites \textbf{(a)} or with site-basis coherence \textbf{(b)}, with corresponding energy-level diagrams (\textbf{(b)} and \textbf{(d)}).
    Donors are weakly coupled to the acceptor, causing the trapping to occur via FRET.
    Trapping rates are proportional to the squares of transition dipole moments (green arrows). 
    In the statistical mixture of site states \textbf{(a)}, the donors transfer excitons to the acceptor independently. In the coherent superposition $\ket{+}$ \textbf{(b)}, the transition dipole moments add constructively, causing an increase in the trapping rate and efficiency.
    The increased dipole moment also leads to faster radiative recombination (orange arrows), meaning that supertransfer is significant only if non-radiative recombination (red arrows) is the dominant loss mechanism.
    }
    \label{fig:super}
\end{figure}

Site-basis coherence (delocalisation) can affect efficiency when trapping or recombination occur through delocalised eigenstates, as opposed to individual sites. These enhancements are possible both through trapping enhancement (type IA) and recombination suppression (type IB), which can be achieved, respectively, through \emph{supertransfer}~\cite{Lloyd2010,Kassal2013,Baghbanzadeh2016,Baghbanzadeh2016a,Giusteri2016,Taylor2018} and \emph{dark state protection}~\cite{Creatore2013,Yamada2015,Fruchtman2016,Hu2018,Rouse2019,Moreno-Cardoner2019}, as we discuss in this section.

Two examples of non-local dissipation commonly arise in excitonic systems. The first is radiative recombination, which is non-local if intra-site distances are smaller than optical wavelengths. The second is long-range F{\"o}rster resonant energy transfer (FRET) to an acceptor, which is a type of non-local trapping.
When the donor-acceptor distance is large (and excitonic coupling weak), the latter occurs directly between energy eigenstates~\cite{Beljonne2009,Baghbanzadeh2016,Taylor2018}.
Both radiative recombination and long-range FRET are mediated by the transition dipole moments between excitonic eigenstates and the ground state, which are given by linear combinations of site dipole moments.
When eigenstate delocalisation is significant, constructive interference of dipoles can give rise to \emph{bright} states, where the overall dipole moment is  enhanced, while destructive interference can give rise to \emph{dark} states, where magnitudes of transition dipole moments are reduced or even zero~\cite{Creatore2013,Yamada2015,Fruchtman2016,Giusteri2016,Higgins2017,Hu2018,Rouse2019,Moreno-Cardoner2019}.

Type IA enhancements use delocalisation to enhance trapping.
In particular, \emph{supertransfer}~\cite{Meier1997,Lloyd2010,Strumpfer2012,Kassal2013,Baghbanzadeh2016,Baghbanzadeh2016a,Giusteri2016,Taylor2018,Taylor2019} uses bright states, whose increased dipole-dipole coupling with acceptor states can accelerate FRET and improve efficiency.
In some cases, donor dipole moments can be arranged so that brighter states are lower in energy, meaning that they do not thermally relax to darker states (Figure~\ref{fig:super}).
For example, supertransfer has been studied in the light-harvesting apparatus of purple bacteria~\cite{Lloyd2010,Kassal2013,Baghbanzadeh2016}, which is characterised by highly symmetric antenna complexes surrounding reaction centres. It has been shown that the symmetric arrangements of dipoles optimise inter-complex transfer efficiency using supertransfer~\cite{Baghbanzadeh2016a}.
Furthermore, it has been shown that inter-complex energy transfer often constitutes the main bottleneck in photosynthetic systems~\cite{Dostal2016}, suggesting that enhancing its rate through supertransfer can have a profound effect on efficiency in these systems.

\begin{figure}[tb]
    \centering
    \includegraphics[width=\linewidth]{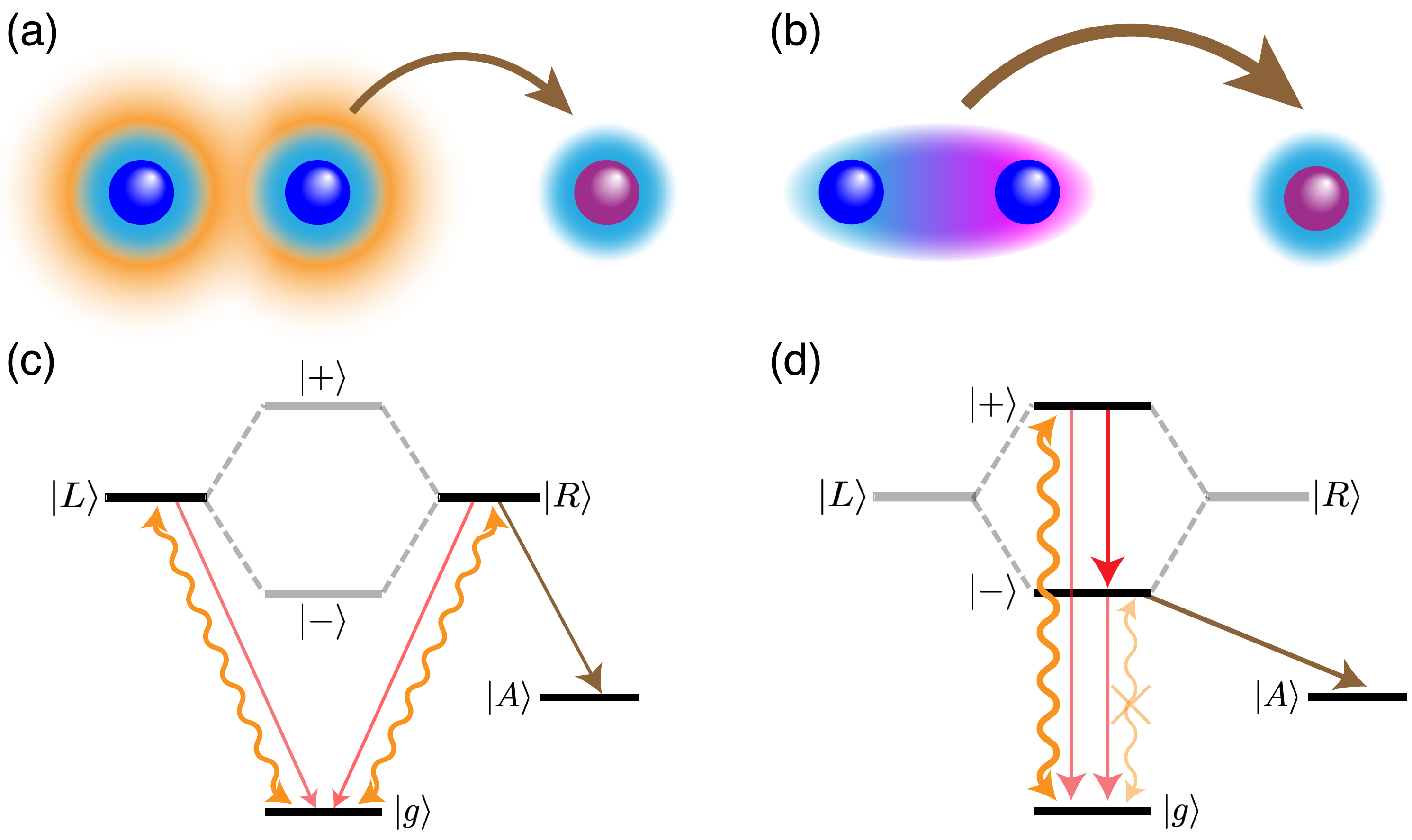}
    \caption{Type IB enhancement via dark-state protection. 
    As in Figure~\ref{fig:super}, a V-type system is coupled to a redshifted acceptor.
    Radiative recombination of excitons (orange glow in \textbf{(a)} and  orange arrows in \textbf{(c)} and \textbf{(d)})---decreases efficiency.
    The energy states $\ket{+}$ and $\ket{-}$ are bright and dark, respectively. Although dark states cannot be optically excited, $\ket{-}$ can be populated through relaxation (red arrows) from $\ket{+}$.
    In the coherent case \textbf{(b)}, the suppression of radiative recombination from the dark state $\ket{-}$ reduces exciton loss and increases the efficiency compared to the statistical mixture of $\ket{L}$ and $\ket{R}$ in \textbf{(a)}. Because dark states are susceptible to non-radiative recombination, this enhancement is only significant if radiative relaxation is the dominant loss mechanism.
    }
    \label{fig:dark}
\end{figure}

In type IB enhancements, delocalisation suppresses recombination. 
It can be realised as \emph{dark state protection}~\cite{Creatore2013,Yamada2015,Fruchtman2016,Higgins2017,Hu2018,Rouse2019,Moreno-Cardoner2019}, where radiative recombination is suppressed by a weakened exciton-radiation coupling in dark states (Figure~\ref{fig:dark}).
While dark states cannot be directly excited by light sources, thermal relaxation from states higher in energy can allow them to be occupied.
If relaxation rates within the excited state subspace are greater than the radiative decay rates of the brighter eigenstates, excitons can be transferred to a dark state before radiative recombination can occur, increasing their overall lifetime.

Dark state protection can only be significant in systems where radiative recombination is the dominant loss mechanism~\cite{Creatore2013,Yamada2015,Fruchtman2016,Higgins2017,Hu2018,Rouse2019,Moreno-Cardoner2019}.
This is not the case, for example, in photosynthetic systems, where recombination is predominantly non-radiative~\cite{Cser2007,Olsen2007,Baghbanzadeh2016,Brumer2018} and mostly unaffected by transition dipole magnitudes.
Furthermore, for dark state protection to work, trapping must not be mediated by eigenstate transition dipoles, such as long-range FRET. Trapping must instead rely on a local transfer mechanism, such as near-field dipole-dipole coupling between sites or via wavefunction overlap~\cite{Creatore2013}.

Supertransfer and dark state protection both take advantage of superpositions of transition dipole moments in low-lying states. However, the former uses constructive interference while the latter uses destructive interference. Therefore, a system cannot use both mechanisms simultaneously.

\begin{figure}[tb]
    \centering
    \includegraphics[width=\linewidth]{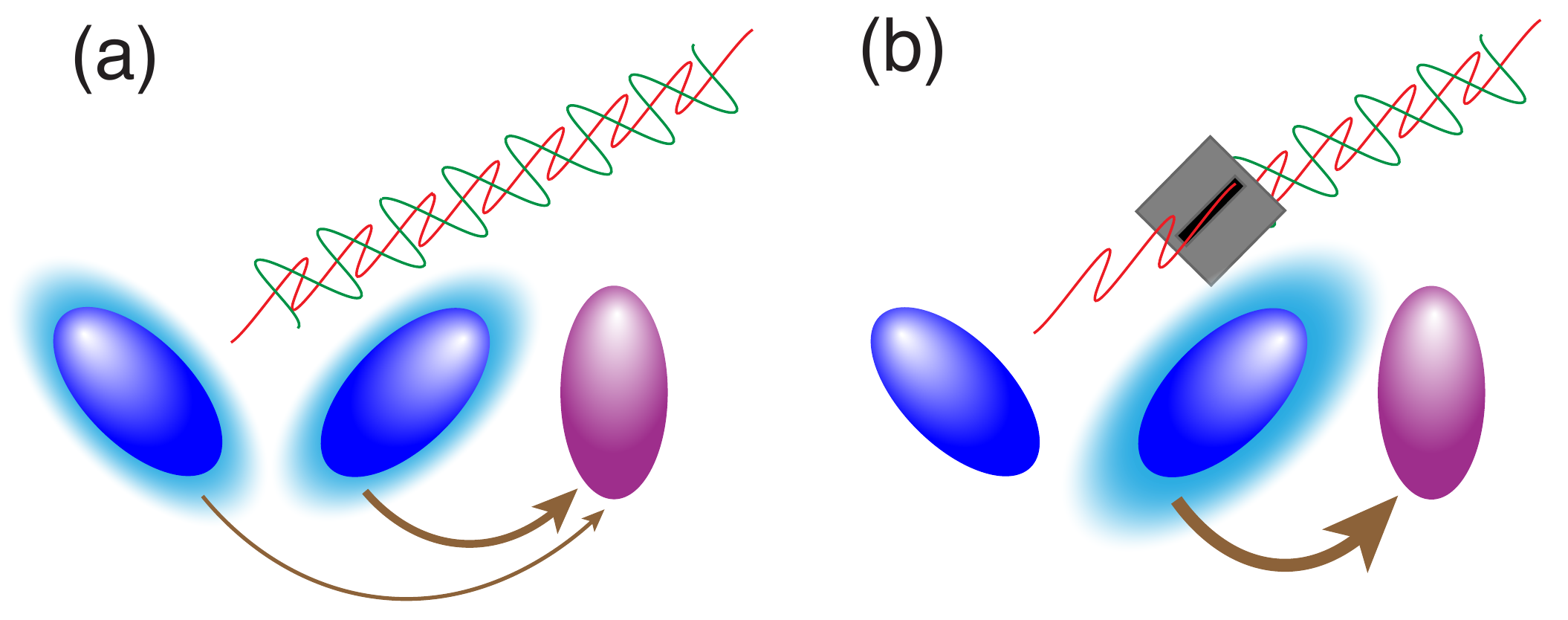}
    \caption{Type IIA enhancement via localisation near trap.
    As in Figure~\ref{fig:dark}, the system contains two donors and a redshifted acceptor, depicted as ellipses to indicate different transition-dipole-moment orientations.
    Eigenstates are delocalised across the two donors and also have perpendicular dipole moments, so that they couple independently to orthogonal light polarisation modes.
    In unpolarised light \textbf{(a)}, the exciton is created in a mixture of the two sites. By controlling the light's polarisation, excitons can be localised close to the acceptor \textbf{(b)}, enhancing efficiency. Adapted with permission from \cite{Tomasi2019}.
    }
    \label{fig:control}
\end{figure}

\section{Type II enhancements}

Energy-basis coherence can affect efficiency when dissipation occurs from individual sites~\cite{Tscherbul2018,Tomasi2019}.
A coherent superposition of eigenstates differs from a mixture with equal populations in that the coherences can, to a greater or lesser extent, localise the exciton on particular sites. As a result, efficiency enhancements are possible if the exciton is localised so that it enhances trapping (type IIA) or so that it reduces recombination (type IIB).

In type IIA enhancement, coherences enhance trapping. 
We previously proposed an example of type IIA enhancement relying on \textit{localisation near the trap}~\cite{Tomasi2019}. The system contains two donor sites coupled to a single acceptor, arranged so that one donor is more strongly coupled to the acceptor than the other, so that trapping occurs almost exclusively from one site (Figure~\ref{fig:control}). Coherence between the delocalised donor eigenstates can localise excitons to either site, either enhancing the trapping (for localisation close to the acceptor) or suppressing it (when localised far from the acceptor).
The scheme allows the coherence to be controlled externally, making the enhancement tuneable. To do so, the donors are arranged with orthogonal dipole moments, resulting in eigenstate dipole moments also being orthogonal.
This allows for the two eigenstates to be addressable independently by two different light polarisation modes and for coherence to be controllable through the polarisation of the light source, allowing for changes in efficiency to be observed due to coherent localisation on either site.

\begin{figure}[tb]
    \centering
    \includegraphics[width=\linewidth]{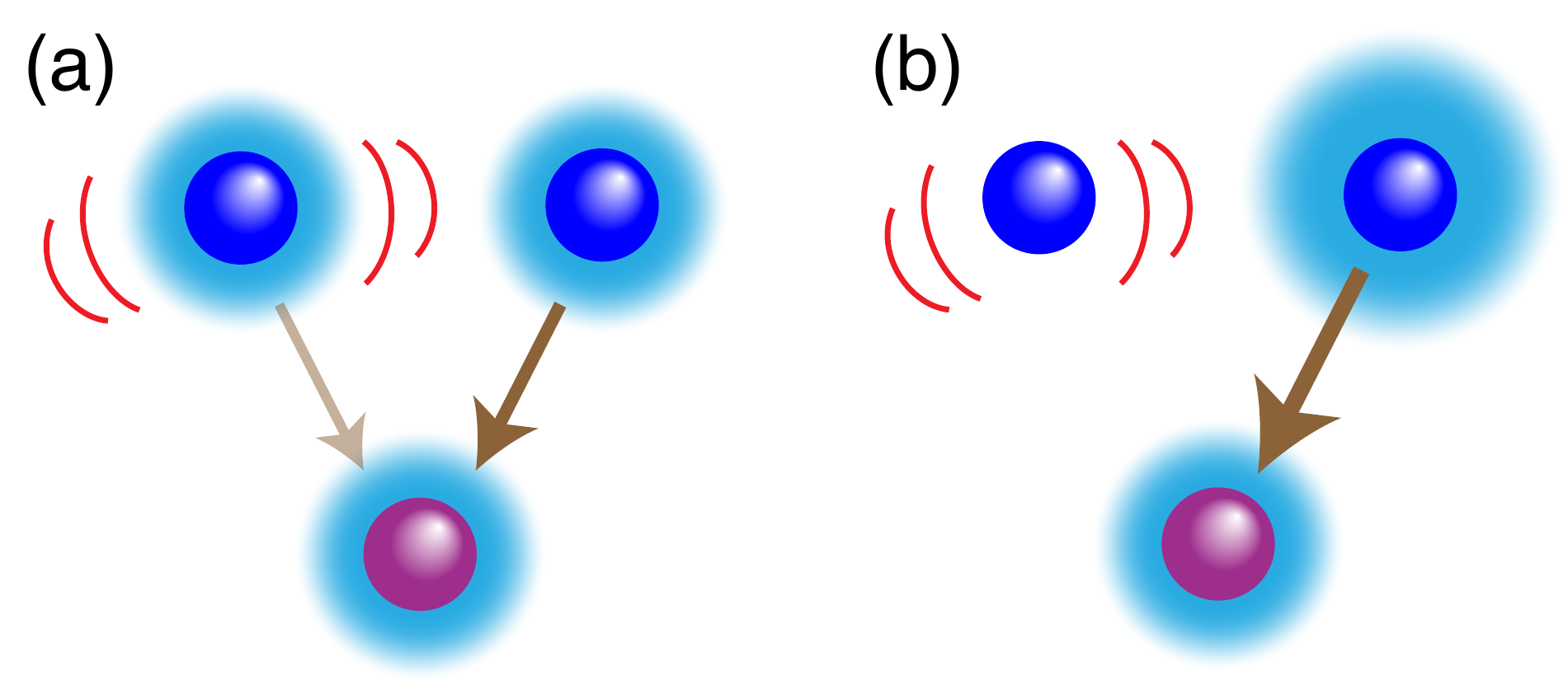}
    \caption{Type IIB enhancement via localisation on quiet site. As in the other Figures, the system contains two donors (blue) and a redshifted acceptor (purple). If one donor experiences strong recombination (red lines), efficiency is lower for an incoherent mixture of both donor sites \textbf{(a)} than when excitons are localised on the less-noisy site through a coherent superposition of eigenstates \textbf{(b)}.}
    \label{fig:noiseloc}
\end{figure}

In type IIB enhancement, coherences suppress recombination. To our knowledge, a type IIB enhancement has not previously been proposed, but it emerges as a possibility from our classification as a process we call \emph{localisation on quiet site}. 
It could be seen in a system that has two donors equally coupled to an acceptor (so that localisation on either site would not affect trapping), but where one donor is affected by stronger local noise, and therefore experiences faster recombination than the other (Figure~\ref{fig:noiseloc}).
In this system, efficiency enhancements would occur if coherence localised excitons on the less noisy site.

For any type II enhancement to be observed, it is not sufficient that excitons are generated in coherent initial states, but also that subsequent oscillations of the coherences are slow enough for the efficiency to be affected.
This oscillation is often rapid compared to dissipative timescales, meaning that, even if coherences were large, their average value would be zero and their influence negligible~\cite{maykuhn,breuer}.

However, if oscillations are slower than the rate at which eigenstate populations are depleted, coherences could have a significant net average value over the lifetime of excited states and therefore influence efficiency.
Energy-basis coherences can therefore only have a significant effect when they are between eigenstates close enough in energy.

\section{Toward experimental demonstrations of enhancements}

Having shown that coherences can enhance efficiency, it is important to consider how these enhancements can occur in the first place and how they can be controlled.
We are not aware of unambiguous experimental demonstrations of any of the mechanisms above, which would be a prerequisite for engineering coherent enhancements into artificial light-harvesting devices. So far, experiments have shown the possibility of engineering site-basis coherence into artificial systems~\cite{Scholes1993,Eisele2012,Rickhaus2017,Bricks2018} and generating long-lived energy-basis coherence in natural~\cite{Engel2007,Liu2007,Panitchayangkoon2010,Collini2010} and artificial~\cite{Hayes2013} systems. However, these experiments did not relate the observed coherences to light-harvesting efficiency, usually because they did not include acceptors to trap the excitons.
Furthermore, experimentally showing coherence-enhanced light-harvesting would require a control without the relevant type of coherence, so that any efficiency improvement could be related to the coherence, and not another variable.

Each of the model systems in Figures~\ref{fig:dark}--\ref{fig:noiseloc} is the simplest possible example of the particular kind of enhancement and could, in principle, be built out of molecules or other nanostructures.
We argue below that, although type I enhancements may be easier to incorporate into real devices, the necessary degree of control required for unambiguous demonstrations may be easier for type II enhancements, which can be controlled externally without introducing confounding variables.

Site-basis coherence needed for type I enhancement characterises delocalised states. These are most easily realised as energy eigenstates that occur if couplings between near-resonant sites are sufficiently strong~\cite{maykuhn,Amerongen,Scholes1993}.
In that case, delocalisation can be generated simply by exciting an eigenstate, either optically or by far-field FRET.
Dark states that cannot be directly excited can sometimes be populated through relaxation from higher-energy eigenstates (Figure~\ref{fig:dark}).

The extent of delocalisation then depends on the eigenstates of the system, which are determined by its molecular Hamiltonian.
Therefore, to control site-basis coherence, one must alter the system's chemical or physical structure~\cite{Scholes1993}.
Unfortunately, structural alteration is likely to introduce confounding variables that can be difficult to control or compensate for~\cite{Baghbanzadeh2016}.
For example, changing the couplings between sites changes not only the extent of eigenstate delocalisation but also their energies, which can significantly change the exciton dynamics~\cite{Baghbanzadeh2016}.

Energy-basis coherence needed for type II enhancements can be prepared optically because light couples directly to eigenstates and not to individual sites~\cite{Jiang1991,Bruggemann2004,Voronine2011,Caruso2012,Hoyer2014, Mancal2010}.
In particular, coherence between eigenstates can be generated when there is coherence between the light modes which excite those eigenstates.
This can be achieved either through coherence between frequency modes~\cite{Jiang1991,Bruggemann2004,Voronine2011,Caruso2012,Hoyer2014,Mancal2010} or, when eigenstates couple to different polarisation modes, using polarised light~\cite{Tscherbul2018,Tomasi2019}.
For example, the energy-basis coherences detected in spectroscopic experiments on light-harvesting systems~\cite{Engel2007,Liu2007,Panitchayangkoon2010,Collini2010,Hayes2013} are due to the spectral coherence of the laser pulses~\cite{Jiang1991,Mancal2010,Brumer2012,Kassal2013,Brumer2018}.

Controlling the exciting light could make it possible to switch coherence on and off, without affecting other system variables, making it a candidate for demonstrations of type II enhancements~\cite{Tomasi2019}.
Indeed, optimal control schemes have been used to control energy transfer dynamics in larger light-harvesting systems, using feedback-controlled pulse shaping~\cite{Bruggemann2004,Voronine2011,Caruso2012,Hoyer2014}, leading to reports of efficiency enhancements~\cite{Herek2002,Savolainen2008}.
However, the optimal pulses that emerge from the feedback-controlled pulse shaping are complicated, making it difficult to determine the states being excited or to attribute the enhancement to a particular coherent mechanism.
Conversely, states in model systems such as those in Figures~\ref{fig:control} and~\ref{fig:noiseloc} are simpler to model, meaning that optical control experiments on these systems could demonstrate type II enhancements more directly~\cite{Tomasi2019}.

\section{Coherent enhancements under incoherent light}

Most light-harvesting systems---from photosynthesis to photovoltaic devices---operate in sunlight. Hence, for coherence-enhanced light harvesting to become technologically relevant, it must take advantage of coherences that can arise from excitation by incoherent thermal radiation.

Type I enhancements can occur in incoherent radiation, since delocalisation can be controlled through the chemical structure of the system alone, without requiring control of a light source. Indeed, as noted above, supertransfer enhances the efficiency of purple-bacterial light harvesting~\cite{Baghbanzadeh2016a} and dark-state protection should also be possible in suitably designed systems with slow non-radiative recombination.

Type II enhancements are more difficult to obtain in sunlight, and are attended with some controversy.
As discussed above, the most effective way to induce energy-basis coherence is through the coherence of the light source~\cite{Jiang1991,Mancal2010}, a method which would not work in incoherent light.
Nevertheless, several studies have proposed that useful energy-basis coherence can be generated under incoherent light in some cases~\cite{Fassioli2012,Tscherbul2014,Olsina2014,Dodin2016,Dodin2016a,Dodin2017,Pachon2017,Tscherbul2018,Brumer2018,Shatokhin2018,Dodin2019}.
We argue below that type II enhancements in incoherent illumination are unlikely, being possible only in limited circumstances.

Incoherent excitation is a stationary process, proceeding through time-independent steady-states that result from the ensemble average of many realisations of the process~\cite{Jesenko2013,Chenu2016}.
It has been argued that even if the ensemble average is incoherent, the energy-basis coherence in each realisation (termed microscopic coherence) may still enhance the efficiency~\cite{Cheng2009}. 
This is not so, because the order in which the expectation value of efficiency and the ensemble averaging are calculated does not change the result~\cite{Kassal2013}:
\begin{equation}
    \mathrm{Tr}({\eta\braket{\rho}_{\mathrm{ens}}}) = \braket{\mathrm{Tr}({\eta\rho})}_{\mathrm{ens}},
\end{equation}
where $\eta$ is the (linear) efficiency operator and $\braket{\cdot}_{\text{ens}}$ represents the ensemble average over all realisations of $\rho$.
Therefore, the average of microscopic efficiencies equals the ensemble-averaged efficiency, meaning that coherences can contribute to the overall efficiency only when they have non-zero ensemble-averaged values.

Energy-basis coherences in incoherent light can have non-zero ensemble-averaged values in two cases: sudden turn on and Fano interference.

The first case, of sudden turn on, is the appearance of time-dependent, transient coherences if incoherent radiation is suddenly turned on~\cite{Fassioli2012,Tscherbul2014,Dodin2016}. A sharp change in excitation conditions implies the process is no longer stationary, and transient changes occur in excitonic states as a result~\cite{Dodin2016,Dodin2016a,Brumer2018}. Essentially, incoherent light that is suddenly turned-on is not actually perfectly incoherent. The induced coherences are transient in the sense that they are short-lived compared to the subsequent duration of the incoherent light. 
Furthermore, if the incoherent radiation is gradually turned on, the magnitude of the transient coherences becomes insignificant even for turn-ons as fast as $\SI{1}{ms}$~\cite{Dodin2016a}. From a practical point of view, transient coherences of this kind are not useful, because a light harvester's performance is judged by its long-time efficiency on physiological timescales.

The second possible source of coherences in incoherent light is Fano interference~\cite{Scully2010,Scully2011,Svidzinsky2011,Dorfman2013,Tscherbul2014,Dodin2016,Dodin2017,Dodin2019}. Broadening of eigenstate energies due to environmental interactions can cause the absorption spectra of two closely spaced eigenstates to overlap, meaning the two eigenstates can be excited simultaneously by the same light mode into a coherent superposition~\cite{Dodin2016}. 
The slow oscillation of coherences within individual realisations---due to the near-degeneracy of eigenstates---means Fano coherences have a non-zero average at steady-state~\cite{Scully2010,Scully2011,Svidzinsky2011,Dorfman2013,Tscherbul2018}.
As a result, Fano coherences can cause type IIA enhancements even under incoherent light, including in systems similar to the example in Figure~\ref{fig:control}~\cite{Tscherbul2018}. As in the discussion above, this effect is limited to close-to-degenerate eigenstates where trapping is faster than the coherent oscillations. 

Several papers reporting large coherent enhancements in incoherent light assume large photon occupation numbers (up to $\bar{n}=\num{90000}$)~\cite{Dorfman2013,Creatore2013}. We caution against this approach. The approximate temperature of sunlight is $\SI{5800}{K}$, giving the average photon number in the visible range of $\bar{n}\approx 0.01$.
Contrary to some suggestions, concentrating the light does not change $\bar{n}$, because lenses only increase the number of modes a system interacts with, not their temperature; a red-hot object does not look white hot in a magnifying glass.

\section{Conclusions}

Whether a type of coherence affects the efficiency of a light-harvesting system depends on whether the trapping and recombination processes affecting it occur via localised or delocalised states.
This observation allows us classify coherent light-harvesting enhancements in type I, due to site-basis coherence, or type II, due to energy-basis coherence, with both further split into sub-types A and B, depending on whether the enhancement results from trapping enhancement and recombination suppression. Overall, our classification may inform the design of experiments to demonstrate the enhancements and of future devices that exploit coherence to harvest natural light. 
In particular, type I enhancements are the more promising candidates for practical harvesting of natural light, even though type II enhancements are likely to be easier to conclusively demonstrate experimentally.

\section{Acknowledgments}

This work was supported by a Westpac Scholars Trust Research Fellowship, by an Australian Government Research Training Program scholarship, and by the University of Sydney Nano Institute through a Grand Challenge and through a scholarship.

\end{document}